# XBadges. Identifying and training soft skills with commercial video games
## Improving persistence, risk taking & spatial reasoning with commercial video games and facial and emotional recognition system

Sergio Alloza[1], Flavio Escribano[1], Sergi Delgado [2,3], Ciprian Corneanu[2,3], Sergio Escalera[2,3.]

[1] Gamification Research Department at GECON.es, c/ Aragó, 336  08009 Barcelona, Spain.
[2] Universitat de Barcelona,  Gran Via de les Corts Catalanes, 585, 08007 Barcelona, Spain.
[3] Computer Vision Center, Campus UAB, Edifici O, s/n, 08193 Cerdanyola del Vallès, Barcelona, Spain.

[1]{salloza, fescribano}@gecon.es, [2,3]sergi1789@gmail.com, [2,3]cipriancorneanu@gmail.com, [2,3]sergio.escalera.guerrero@gmail.com

**Abstract.** XBadges is a research project based on the hypothesis that commercial video games (nonserious games) can train soft skills. We measure persistence, spatial reasoning and risk taking before and after subjects participate in controlled game playing sessions. In addition, we have developed an automatic facial expression recognition system capable of inferring their emotions while playing, allowing us to study the role of emotions in soft skills acquisition. We have used Flappy Bird, Pacman and Tetris for assessing changes in persistence, risk taking and spatial reasoning respectively. Results show how playing Tetris significantly improves spatial reasoning and how playing Pacman significantly improves prudence in certain areas of behavior. As for emotions, they reveal that being concentrated helps to improve performance and skills acquisition. Frustration is also shown as a key element. With the results obtained we are able to glimpse multiple applications in areas which need soft skills development.

## 1   Introduction

There is no consensus about what soft skills are and an absolute description or classification is missing. Among the various perspectives taken by different authors, one of the most useful is defining soft skills in relation to the workplace. In this sense, soft skills can be seen as interpersonal, human, people or behavioral skills necessary for applying technical skills and knowledge in the workplace [1] or as a new way to describe a set of abilities or talents that an individual can bring to the workplace [2]. Additionally, soft skills can be categorised as: (1) intrapersonal and interpersonal skills; (2) personal and social skills; and (3) cognitive skills [3] and can also be attributed intensity levels [4, 5].

In this context, it is very important to measure and boost soft skills. Some authors think that [6, 7, 8] as organizations become increasingly diverse, the ability to exhibit some soft skills like critical thinking or decision making with confidence can provide greater opportunities for employment. Unfortunately, even though communication and soft skills are noted by employers as important skills in the workforce, they are highly lacked by recent graduates applying for employment [9, 10, 11].

Employers are looking for methods to reduce the costs of identifying soft skills through behavioral interviews. Unfortunately, such procedures are subjective, expensive and time-consuming. Furthermore, they cannot be used to filter out large amounts of CVs in the initial stages of the hiring process [12].

In the world of formal education the landscape of identification and evaluation of transversal competences is perhaps even more complex than in the labor market since these skills are hardly being evaluated and trained. In most cases, there is no standard and quantitative system available for teachers. Transversal competencies hardly get evaluated and trained. As a consequence, it is difficult to add soft skills related training as this would overload the teaching agenda and would entail additional costs.

In this context, video games are very useful identification and training tools. Nowadays, they are a very popular element, in fact they have become the most used artifact in the entertainment industry. In addition their presence begins to influence other aspects or areas a priori non-ludic.

This study was born from a research project called XBadges, cofunded by the Ministry of Industry, Energy and Tourism, Government of Spain, 8th in the AEESD call, forming a consortium led by the company COMPARTIA, which subcontracted both GECON.es foundation (gamification experts) and the University of Barcelona (for the creation of an artificial vision system).

The objective of XBadges is to facilitate, with the use of software, the training and evaluation of soft skills of the users through the use of commercial video games and, if required, to grant certifications of the acquisition. Thus, with XBagdes both the business and education sector would have a tool to meet their current and future needs related to the identification and training of soft skills. Much of the current literature is already investigating the effects of video games on human cognition [13, 14, 15], but very few studies [16, 17, 18, 6] relate these to soft skills.

Specifically, and following an initial review of commercial video games in open source and preliminary soft skills (commented below), the following games and soft skills were chosen as hypotheses for the research: Pacman (Risk Taking), Tetris (Spatial reasoning) and Flappy Bird (Persistence). These soft skills were chosen based on a review of the competencies most valued by various organizations and institutions [19, 20].

As for the video games, open source video games were needed to embed internal indicators within the code and measure the skills with our telemetry algorithm (actions done by the player). That telemetry was created thanks to a literature review specific to each skill, as we show next. These review was also another reason why we have chosen the mentioned games, since they have the necessary elements to stimulate the skills but they are still technically simpler than most of the current commercial video games (so we can link a specific behavior and embed the code to track skill acquisition into the game).

For tracking persistence with Flappy Bird, as [21, 22] said: "A subject is persistent when, faced with a situation in which it has to emit responses to reach a given solution (reach a score of 20 in Flappy Bird, for example), it maintains a high response rate (the user keeps trying) despite the low frequency of reinforcement (the user keeps dying)". So, in this case, the telemetry was tracking the tries over the time and how far the subjects reach.

In Pacman the telemetry tracks the behaviors that are risky, like being near a ghost, eliminate them when they are vulnerable but the player knows there is only a little time of invulnerability left, etc. So following [23, 24] we can infer that the behavior behind these tracked actions is Risk Taking (derived from decision making).

Finally in the case of Tetris, we decided to replicate -as control measure- some hypothesis [25] since there are already a lot of research about how Tetris can change our minds [26, 27, 28] and study [25] specifically measure Spatial Reasoning. The telemetry in this game tracked completed lines and the time between pieces placement, based on the premise that repeated exposure to changing visual patterns in a 2D virtual space (manipulable under rotation and translation) with progressive difficulty increase curve will decrease the time required for information processing, when processing speed being faster and rotation and translation more effective (set of skills that combine spatial reasoning) as the previous authors argued.

The software also captured the emotions thanks to the system of artificial vision. Once a face is detected, emotion recognition is performed in the corresponding bounding box/area of interest. For the recognition of emotions in images, we based on deep learning, in particular we took benefit of the pre-trained VGG convolutional neural network to be fine-tuned on emotions considering annotating public emotion datasets. As a result, a deep learning model was trained, able to recognize face textures representative of the presence of a particular emotion. Thus allow us to see the effect of the emotions of the users in the data and in the competences acquisition.

The emotions we have captured are of Joy, Frustration, Concentration and Boredom. The selection of these emotions has been made taking into account the most studied emotions and the research behind the recognition of emotions [29, 30].

The objective of the research is to contrast the following hypotheses:

1. Commercial video games as a pedagogical differentiator elements

(nonserious games), improve soft skills.

2. The percentage of emotions generated at a general level correlates with the percentages of improvement of the users.

3. Emotions generated by users at specific times vary according to the scores obtained from some indicators (completing a line in Tetris and eliminating ghosts in vulnerability mode A or B in Pacman).

## 2 Methods

### 2.1 Participants

The sample consists of 15 subjects (12 males and 3 females), randomly divided into 3 groups of 5 people each (group 1 to FlappyBird, group 2 to Tetris and group 3 to Pacman). We have chosen 5 as minimum number of people for statistical analyzes to be reliable and valid, as indicated in [31] and taking as reference other studies that also have a reduced sample size [32, 33]. The inclusion criteria applied in the sampling is:

- Age between 18 and 50 years.

- Not accustomed to playing the video games of the research or similar (minimum 1 year without previous experience).

The participants were searched through the social networks, personal contacts and also thanks to a collaboration with Yuzz Sant Feliu (center of innovation and co-working).

### 2.2 Materials

On one hand, one of the materials we have is the XBadges software platform. This software has integrated the three video games previously commented with the added telemetry. It also has the aforementioned system of facial and emotional recognition that allows the recording of the emotions of the players while playing, as we will explain next.

On the other hand, we list below the standardized tests that have been used as a reference measure to test whether or not there is a real acquisition of the mentioned soft skills and thus to verify the validity of the indicators as measuring instruments:

- **Risk Taking - Domain-Specific Risk-Taking (Pacman)**

Domain-Specific Risk-Taking (DOSPERT) [34] is a psychometric scale that assesses risk taking in 5 different domains: financial decisions, health/safety, recreation, ethics, and social decisions. The subjects rate the likelihood of specific risk activities for each domain. A second and third part of the questionnaire assesses the perception of the risk magnitude of the expected benefits of the activities of the 1st part. The reduced and revised spanish version of 30 items [35] is used.

- **Persistence - Big Five subscale (Flappy Bird)**

As a personality test, the Big Five Questionnaire allows us to observe patterns and profiles of behavior in users. It has multiple questions grouped in different dimensions. Specifically, the dimension "Conscientiousness", which bifurcates in two sub-dimensions: "Scrupulousness" and "Perseverance". Given the purpose of the experiment, we are interested only in the subscale that measures Perseverance (Persistence). Again a spanish version is used [36].

- **Spatial Reasoning – Fibonicci's Test (Tetris)**

We have used the web test [37] that was used in the study [25] to measure spatial reasoning ability. This test consists of a series of items showing a series of 3D figures displayed and the subject has to choose one option (between 4) of the same figure, but folded.

### 2.2.1 Automatic Facial Expression Recognition

Facial expressions are strong predictors of affective states. In order to automatically infer the affective state of subjects while playing video games we have built an Artificial Vision System (AVS) capable of recognizing a set of predefined expressions from facial images.

Using human annotators to manually label facial images with one of the predefined expressions is a cumbersome process, prone to subjectivity and human errors. Being able to train an automatic prediction model, it opens the way to detecting facial expressions in large amounts of data in an objective way allowing extended statistical analysis.

In order to detect the four predefined emotions on the face (Joy, Concentration, Frustration and Boredom) we have mapped each emotion into a corresponding universal facial expression. In this sense we use Neutral faces as marker of Concentration, Happy faces as marker of Joy, Sad faces as marker of Boredom and both Angry and Disgusted faces as markers of Frustration.

**Model**. We have train a deep neural network architecture in order to classify a face into one of the targeted facial expressions. The network follows GoogleNet [38], a well known architecture in the machine learning community which has been successfully used for many visual pattern recognition tasks. The network is based in the repetition of the same module (called Inception) in a stacked manner, following the idea of network in network. This module is repeated nine times inside GoogleNet and is composed by a first level of 1x1 convolutions and a 3x3 max pooling and a second level of 1x1, 3x3, 5x5 convolutions. After each Inception module, there is a filter concatenation step that joins all previous results. The width of Inception modules ranges from 256 filters (in early modules) to 1024 in top Inception modules. Given the depth, propagating gradients back through the network is problematic. In order to alleviate the vanishing gradient problem 2 auxiliary classifiers are connected to intermediate layers of the network. At inference time, these auxiliary networks are discarded.

**Pretraining**. Due to its large number of parameters, training GoogleNet from scratch would have required large amounts of data. As our facial expression dataset is relatively small we use transfer learning for initializing the network's weights. The initial network weights were found by training the network for Age/Gender facial classification using hundreds of thousands of images, coming from a filtered mix of Imdb-Wiki [39] and Adience [40] datasets. This previous network is specialized to detect details in faces and was a good initial point for facial expression classification.

**Training**. Public data containing universal facial expressions of emotion is widely available, making possible the training of complex models capable of learning statistical relations between the morphology of the face and target classes. In "Fig. 1", a selected set of examples used to train our model are depicted.

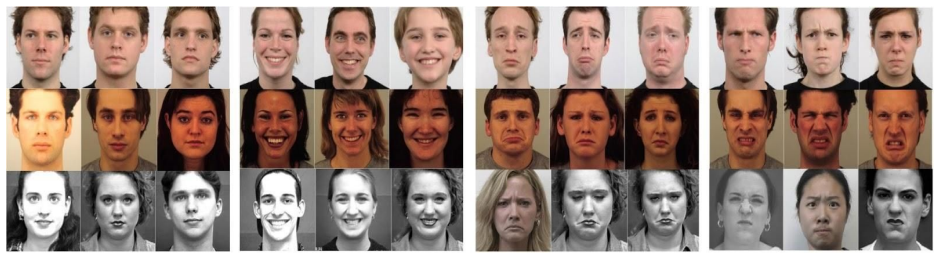

**Fig. 1.** Examples of targeted facial expressions during training. From left to right: Neutral, Happy, Sad and Angry. The training samples were collected from 3 different datasets. On top row Radboud [41], middle row KDEF [42] and on the bottom CK+ [43].

The data used is a compilation of three facial expression datasets, totalling 6562 images. About 55% is from the Cohn-Kanade dataset [43], 15% from the Radboud dataset [41] and approx. 8% from KDEF dataset [42]. In order to keep class balance, additional 1461 images (about 22% of the total data corpus) were collected from queries on the web.

During fine-tuning, 75% of the dataset is used for training and the rest of 25% for validation. The learning rate has been set to 0.01 with an automatic decrease of 1/10 every 33% of the training phase. The Stochastic Gradient Descent (SGD) optimization method was used with a batch size of 32 images. We have stopped the fine tuning after 100 epochs, at this stage the network performance begin saturated.

## 2.3 Procedure

In the first place, the subjects completed the questionnaires corresponding to each video game (as a pre-test phase), explained in the previous section. Then each participant played the video game that corresponds to his group for 3 sessions of 40 minutes each, sessions distributed at the convenience of the subjects in a maximum period of 1 week, with no possibility of doing two or three sessions in one day. The design of temporality is based on a similar study [28]. In the Flappy Bird group the subjects were able to stop playing whenever they wanted after the 20th minute. When measuring Persistence, we had to leave a margin of time in which the subject decided to play or not, since otherwise we would have been skewing the persistence scores. Finally, at the end of the 3rd session, the subjects had to re-complete the questionnaires (as a post-test phase) so we were able to compare the questionnaire results before and after the game training.

# 3 Results

After the analysis of the data obtained, the following results are presented, grouped by hypotheses:

## 3.1 1º Hypothesis: Video games & soft skills.

The data are shown in table form, assembled by indicators / telemetry and standardized tests, by videogame:

- *Flappy Bird*

Sessions data have been modified by removing the high end values (40 minutes) for the "ceiling effect". Some of the data (about 30%) have been provoked by the end of session time, so we cannot infer that these are the times users would adjust if they had more time to play. The final table after clustering and descriptive analysis of the data is as follows "Table 1":

Table 1. Total time indicator data of Flappy Bird.

|  | *1º Session* | *2º Session* |
|---|---|---|
| Average | 31' | 38' |
| Standard error | 2,415' | 0,913' |

Statistical significance of Student's t-test for paired samples accepted (t= -2,818, p=

0,033). Below are the data obtained through telemetry in video games "Table 2". This data have been cleaned of registry errors:

**Table 2.** Telemetry data of Flappy Bird.

|                | 50'   | 60'   | 70'   | 80'   |
|----------------|-------|-------|-------|-------|
| Average        | 0,082 | 0,092 | 0,111 | 0,116 |
| Standard error | 0,045 | 0,049 | 0,059 | 0,058 |
| *90'*          | *100'*| *110'*| *120'*| *130'*|
| 0,120          | 0,124 | 0,130 | 0,138 | 0,147 |
| 0,056          | 0,057 | 0,060 | 0,062 | 0,062 |

Statistical significance of repeated measures ANOVA accepted with epsilon GG adjustment (Epsilon GG= 0,17; F= 10,003, p= 0,025). The results of the standardized test Persistence Big Five subscale "Table 3" are presented, with statistical significance of Student's t-test for paired samples not accepted (t= -1,176, p= 0,152):

**Table 3**. Data of Perseverance Subscale of Big five.

|                | Pre phase | Post phase |
|----------------|-----------|------------|
| Average        | 47        | 48,8       |
| Standard error | 1,923     | 1,827      |

- *Tetris*

Data obtained from the Tetris telemetry "Table 4":

**Table 4**. Telemetry data of Tetris.

|                | 10'     | 70'     | 80'     |
|----------------|---------|---------|---------|
| Average        | 681,200 | 798,836 | 789,279 |
| Standard error | 122,486 | 96,582  | 113,516 |
| *90'*          | *110'*  | *120'*  | *130'*  |
| 755,369        | 773,241 | 779,983 | 792,213 |
| 90,746         | 107,316 | 115,917 | 113,825 |

Statistical significance of repeated measures ANOVA accepted with epsilon GG adjustment (Epsilon GG= 0,35, F= 6,614, p= 0,020). The results of the standardized test that measures spatial reasoning are shown below "Table 5" with statistical significance of Student's t-test for paired samples accepted (t= -2,449, p= 0,035):

**Table 5**. Spatial reasoning test results.

|  | *Pre phase* | *Post phase* |
|---|---|---|
| Average | 12,4 | 14,2 |
| Standard error | 1,631 | 1,772 |

- *Pacman*

Data obtained from the Pacman telemetry "Table 6 ":

**Table 6**. Telemetry data of Pacman.

|  | *10′* | *20′* | *30′* | *50′* |
|---|---|---|---|---|
| Average | 48,930 | 49,008 | 54,733 | 53,199 |
| Standard error | 9,350 | 10,094 | 9,551 | 10,128 |
| *90′* | *110′* | *120′* | *130′* | |
| 755,369 | 773,241 | 779,983 | 792,213 | |
| 90,746 | 107,316 | 115,917 | 113,825 | |

Statistical significance test of repeated measures ANOVA not accepted with epsilon GG adjustment (Epsilon GG= 0,22; F= 2,499, p= 0,170).

Regarding the results of DOSPERT (test that measures Risk Taking), no statistically significant differences were found in general or in any of the subscales except for Safety and Health (only in the part of the test that measures probability of behavior), where a significant difference of means was found through the Student's t-test for paired samples (mean pre= 22,6 & mean post= 19; t= 2,882, p= 0,022).

### 3.2  2º Hypothesis: Emotions & improvement percentage.

The correlations between the four emotions (J= Joy, C= Concentration, F= Frustration and B= Boredom) and the percentage of improvement of the three video game

indicators (FB= Flappy Bird, T= Tetris and P= Pacman) are shown "Table 7":

**Table 7**. Pearson's correlations (r) and Spearman's (rho) between emotions and improvement percentage. *Statistically significant correlation (p= 0,02).

|    | J        | C       | F        | B        |
|----|----------|---------|----------|----------|
| FB | r= -0,86 | r= -0,85| rho= 0   | r= 0,49  |
| T  | rho= 0,3 | r= 0,66 | r= -0,53 | rho= 0,5 |
| P  | r= 0,16  | r= 0,93*| r= -0,77 | r= -0,41 |

### 3.3  3º Hypothesis: Emotions & video game indicators.

Next, the averages of the percentages of emotions present in the moments in which the indicated criteria were fulfilled are presented "Table 8", following the abbreviations of the previous hypothesis. The indicators are: completing a line in Tetris (reflected in the table as Tetris) and eliminating a ghost in vulnerability mode A and B in Pacman (reflected as Pacman 1 and Pacman 2 in the table):

**Table 8**. Average of emotions percentages present in each indicator.

| *Indicators* | J %   | C %    | F %    | B %   |
|--------------|-------|--------|--------|-------|
| Tetris       | 7,56% | 50,32% | 34,21% | 7,85% |
| Pacman 1     | 5,65% | 25,36% | 63,61% | 5,37% |
| Pacman 2     | 4,29% | 25,23% | 65,46% | 5,07% |

## 4  Discussion

Interpreting the results, we can affirm the following premises, again by each hypothesis:

### 4.1  1º Hypothesis: Video games & soft skills.

- *Flappy Bird*

Significant differences were detected in the data obtained through the Flappy Bird indicators, with a total time of 130 minutes of training (F= 10,003; p= 0,025). However, we cannot say that these changes reflect an improvement in persistence capacity outside Flappy Bird, given the nonsignificance changes in the Big Five subscale measures (t= -1,1766; p= 0,152).

In spite of this we can establish new lines of investigation guiding the video game Flappy Bird as a measure of persistence more sensitive than the standardized test itself, since although the change in the questionnaire is not significant, the average of the scores of the same one rises (47 vs 48,8). Under this line we would face a nonsignificance caused by a small sample size, memory bias when repeating the same test in just 1 week, little training time or any other variable outside the game. Supporting this new hypothesis, one of the indicators that theoretically relates more to Persistence, "playing time, number of tries" (response frequency), did show significant changes (t= -2,818; p= 0,033) indicating that players spent more time in the game the longer they played.

- *Pacman*

As can be seen previously, significant changes have been detected in one of the DOSPERT subscales. In particular, against the approach of the initial hypothesis, there is a decrease in the probability of risky behavior in the area of Health and Safety (t= 2,882; p= 0,022), so we can say that playing Pacman with a training time of at least 90 minutes, increases the prudence in the mentioned area.

This is an unexpected result since we believed Pacman would increase the risk taking behaviour instead of diminish it. We believe that this reduction in risk-taking behavior may reflect an adaptation to the game strategy. As the subject plays Pacman, it is more aware of the risks that exist within the game and adjusts its strategy to get more points and die less, reducing risky behaviors.

It is also pleasantly surprising that the behaviour change in the video game may reflect a change in the actual risk taking behavior. With these results several applications could already be seen, for example in the clinical field where impulsivity or recklessness are very present in most mental disorders.

- *Tetris*

Tetris training with a minimum of 70 minutes of play has been shown to significantly improve spatial reasoning ability (F= 6,61449348; p= 0,02). These results fit the replica of the study [25] where they also relate the same video game and spatial reasoning, obtaining similar results. In addition this research also specifies the improvement effect of Tetris since the sessions have been carried out with a lower sample size in regard to the original study.

We also emphasize in a general way, that not having measured other soft skills, we are leaving aside relations that can be significant. A good way to evolve the research would be to expand the range of capabilities to measure and relate them to different (or the same) video games.

We also discuss the limitations of the memory effect in the tests complementation of the post phase (pre-post test design), the small sample size per group and the short temporal design of experimental sessions, so that the results obtained could be underestimated (statistical error Type II).

### 4.2  2° Hypothesis: Emotions & improvement percentage.

The results obtained regarding the emotions related to the percentage of improvement of the indicators, do not follow the initial approach. In fact, only one correlation of the 12 (4 emotions * 3 games), Concentration & Pacman, is significant (r= 0,93, p= 0,02), showing that the more concentrated Pacman is played, more is the improvement playing the video game.

The value of the correlation is very sensitive to the number of data available to analyze, so if the study had been carried out with a large number of people and therefore, there would be many more data to analyze, the value of the correlations would oscillate as well as their statistical significance, confirming perhaps the initial hypotheses that to more presence of boredom less percentage of improvement, or greater the presence of joy is, greater the percentage of improvement, for example.

### 4.3  3º Hypothesis: Emotions & video game indicators.

To our surprise, joy was not one of the most prevalent emotions when these indicators were met. In particular, in Tetris, when completing lines during the games, the subjects showed high concentration percentages (50,30%) while the other emotions did not have as much presence. An example of what is commented "Fig. 2". In Pacman, while reaching and eliminating ghosts in vulnerability mode A and B, the prevailing emotion was in both cases frustration (with more than 63% in both cases).

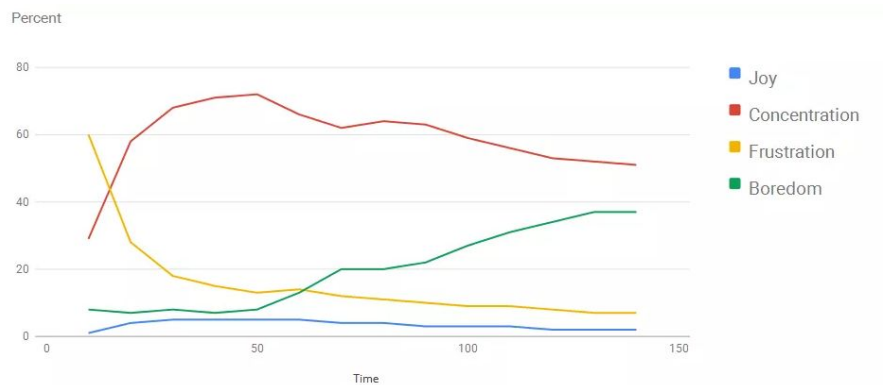

**Fig. 2.**   Artificial vision module inside XBadges platform. Distribution example of emotions of a Tetris player.

Contrary to expectations, concentration and frustration are present in moments where the user is positively reinforced by the video game. Perhaps we are facing here an implicit relationship between these emotions and the acquisition of skills.

The analysis and global interpretation of the results suggest that video games can be useful tools to enhance or boost certain soft skills, as well as the presence of emotions is closely linked to the motivation of the players and their development of soft skills within the game. With this findings, the commercial video games (not only serious ones) win value as a training tool for soft skills, offering them as a new form of tool for markets with possibility of application in multiple sectors.

In the business world, on one hand, providing employees with a tool for identifying and training soft skills required in certain jobs, and on the other hand, to employers, offering a more automated CV screening tool.

As for the academic world, showing the use of video games as a methodology to enhance soft skills which remain unrecognized in most academic curriculum and thus better prepare students to adapt to the context that awaits them.

Another sector where XBadges idea could be applied is eHeatlh. There are many diseases or pathologies that impair certain soft skills. Although, in particular, more research is needed in this field, alleviating certain symptoms or improving dysfunctional skills with video games could prove to be an effective method in addition to engage to the patient.

And above all, regardless of the sector of application, XBadges offers information to the population about the positive influence of their play habits on their minds and behavior, since players will continue to play the same, but knowing that they are boosting their abilities.

**Acknowledgments**. We thank Mercè Muntada (mmuntada@compartia.net) for her support as the leader of Compartia & leader of the project consortium.